\documentclass[aps,prl,reprint,groupedaddress,showpacs,amsmath,amssymb]{revtex4-1}

\usepackage{graphicx, comment}
\usepackage{dcolumn}
\usepackage{bm}
\usepackage{textcomp}

\begin{document}


\title{Nonlinear plasmonic switching in graphene stub nanoresonator loaded with core-shell quantum dot}


\author{M.Yu. Gubin}
\affiliation{Department of Physics and Applied Mathematics,
Vladimir State University named after A. G. and N. G. Stoletovs, Gorky str. 87, Vladimir,
600000, Russia}
\author{A.Yu. Leksin}
\affiliation{Department of Physics and Applied Mathematics,
Vladimir State University named after A. G. and N. G. Stoletovs, Gorky str. 87, Vladimir,
600000, Russia}
\author{A.V. Shesterikov}
\affiliation{Department of Physics and Applied Mathematics,
Vladimir State University named after A. G. and N. G. Stoletovs, Gorky str. 87, Vladimir,
600000, Russia}
\author{A.V. Prokhorov}
\email{avprokhorov33@mail.ru}
\affiliation{Department of Physics and
Applied Mathematics, Vladimir State University named after A. G. and N. G. Stoletovs, Gorky str. 87, Vladimir,
600000, Russia}



\begin{abstract}
We study theoretically a problem of nonlinear interaction between two surface plasmon-polariton (SPP) modes propagating along the graphene waveguide integrated with a stub nanoresonator loaded with a core-shell quantum dot (QD). The conditions of strong coupling and very narrowband resonance for Ladder-type SPP-QD interaction in nanoresonator leading to $\pi$-phase shift of signal SPP mode are established and discussed. Using the full-wave electromagnetic simulation we demonstrate that turning on the pump SPP leads to a transition from destructive to constructive interference in the stub nanoresonator and subsequent effect of switching from locking to transmitting the signal SPP through the waveguide. On the base of the obtained results we develop the model of all-plasmonic graphene transistor with optimized characteristics including the size about $25$ nm and the clock frequency about $100$ GHz functioning at the near-infrared wavelength range.
\end{abstract}


\maketitle

\section{Introduction}
The achievements of modern 2D material science~\cite{ali,nov1,nov2,nov3}, graphene nanotechnologies~\cite{Freitag} and quantum nanophotonics~\cite{boj1} give hope to the experimental implementation of a new information processing gate with record values of transistor size and clock frequency in the soon time. The major premise of this implementation is the new methods of single photon states control~\cite{Luk} and manipulation of single plasmons~\cite{bit} and surface plasmon-polaritons (SPPs)~\cite{Vasa,Stebunov3} under strong-coupling condition~\cite{garc}. Moreover, the control of plasmonic resonance for nanostructures placed near a graphene layer~\cite{Chen} in condition of strong coupling~\cite{garc} may provide the realization of complete quantum computing protocols~\cite{kulik} at a subwavelength scale. At the same time, the expected achieving of high-temperature superconductivity in graphene and doped graphene~\cite{wang,sup1} should solve the key problem of huge losses, which restricts the implementation of plasmonic nanostructures~\cite{wu} in computer technologies. In this context, plasmonic information processing circuits using graphene-localized SPPs as information carriers and quantum dots (QDs)~\cite{fedor} as information processing centers can be considered as good candidates for replacing of existing electronic circuits.

However, even under ideal conditions of strong coupling and ultra-low losses, the problem of very broadband plasmon-exciton resonance remains relevant for semiconductor quantum dots. The solution of this problem may consist in the using narrowband resonances~\cite{evl1} and positive feedback of nanoresonator~\cite{cao}. The stub resonator~\cite{stub1} used for broadband signal filtering can be chosen as a technological platform for this solution. Such resonators are usually made from a combination of metal and dielectric. However, application of graphene~\cite{Liu} and nanostructured graphene resonators~\cite{kong} loaded with 0-D chromophores can help to implement the well-known effects of matter-field interactions such as nonlinear wave mixing~\cite{Cox}, Fano resonance~\cite{kiv,bon2}, slow light~\cite{Kim} and others~\cite{brit,Stebunov1} for creating of plasmonic graphene circuits at infrared radiation (IR) range. Additional mode selections and enhance in the efficiency of plasmon-exciton interactions in such systems can be obtained by using chiral QDs~\cite{Puri}.

\begin{figure}[t]
\includegraphics[width=\columnwidth]{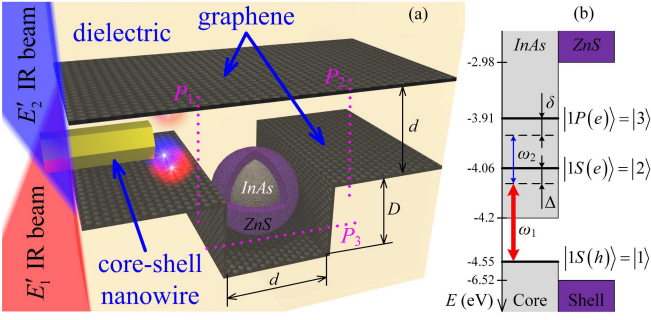}
\caption{\label{fig:1} (a) The model of stub nanoresonator based on the pair of graphene sheets and core-shell QD; (b) the Ladder-type interaction scheme of two SPP modes $E_{1}$ (pump) and $E_{2}$ (signal) and InAs/ZnS QD with energy levels $E_{\left|1\right\rangle} =-4.55 \; \textrm{eV}$, $E_{\left|2\right\rangle} =-4.063 \; \textrm{eV}$, $E_{\left|3\right\rangle} =-3.908 \; \textrm{eV}$ for core radius $a_{QD} =9.9 \; \textrm{nm}$. Bands positions in the bulk: $E_{V1} =-4.55 \; \textrm{eV}$ for top of the valence band and $E_{C1} =-4.2 \; \textrm{eV}$ for bottom of the conduction band in InAs; $E_{V2} =-6.52 \; \textrm{eV}$, $E_{C2} =-2.98 \; \textrm{eV}$ the same in ZnS~\cite{plazm}.}
\end{figure}

In presented work, we carry out a fundamental research of interaction between graphene-like materials and semiconductor quantum dots. Following to Ref.~\citenum{stock}, we propose to integrate a stub nanoresonator with the graphene waveguide, and to load a core-shell quantum dot (QD) into stub for the realization of plasmon-exciton resonance, see Fig.~\ref{fig:1}. In this configuration, the combination of the nanoresonator with the waveguide can lead to a significant narrowing of plasmon-exciton resonance spectral line and precise tuning of the system to the working wavelength.
Moreover, we choose an additional strong pump SPP mode for the realization of controllable switching between two stable states of the signal SPP mode using three-level Ladder-type scheme of SPP-QD interaction by analogy with optics~\cite{Luk1,our2}. The key idea of our work is to use a strong nonlinear interaction between the signal and pump SPPs in a nanoresonator loaded with QD to control the phase shift and transmittance of signal SPP through the stub due to the changing of pump SPP intensity and its detuning. This is necessary to achieve $\pi$ radians phase shift of signal SPP, under which a change in the SPP pattern from destructive to constructive interference in nanoresonator occurs.
The use of a core-shell QD is aimed on decreasing the nonradiative relaxation rate~\cite{reiss} and solving the blinking problem~\cite{naum}. Thus, the main goal of this paper is to investigate and optimize the material parameters and geometric characteristics of nanoresonator loaded with QD in order to get the all-plasmonic gates at the nanoscale for IR range. To accomplish this goal, we will adhere to the semiclassical approximation for the description of nonlinear plasmon-exciton interactions, as well as the strong coupling condition to obtain the steady-state solutions for signal SPP in both basic states of device. The necessary technical equipment and the applicability of various process-technologies for the practical creation of the described devices are discussed.

In the technical framework, the presented stub nanoresonator loaded with QD can be used for practical implementation of all-plasmonic transistors. The use of graphene as a basis for such devices, potentially, allows us to work beyond adiabatically approximation and manipulate with ultrafast electromagnetic fields~\cite{Kelardeh}.
Such devices can be integrated into plasmonic circuits~\cite{evl3} supported terahertz clock frequency and the nanometer gate scale.

\section{Results and discussion}
Let us to consider the system of graphene waveguide based on the pair of graphene sheets and graphene stub nanoresonator loaded with a core-shell quantum dot, see. Fig.~\ref{fig:1}a. The IR fields $E_{1}^{\prime}$ and $E_{2}^{\prime}$ convert into SPP modes $E_{1}$ and $E_{2}$, respectively, propagating in waveguide at the same frequencies than IR fields. We assume that a signal SPP mode $E_{2}$ with frequency $\omega_{2}$ inputs into the system through Port 1. From the experimental point of view, SPPs could be generated in graphene waveguide using special tips of a near-field optical microscope simulating of both either electric dipole (ED) or magnetic dipole (MD) sources~\cite{Picardi,Feber}. However, this approach is good only for a laboratory experiment, since the tip sizes are still large comparing with the scales of the designed plasmonic gates. Therefore, it is necessary to integrate core-shell nanoparticles~\cite{Arnold} or nanowires~\cite{Ho} into the circuit to use them as sources of near field, see Fig.~\ref{fig:1}a. In our simulation we use MD source.

During the propagation SPP mode interacts with QD in Port 3 and leaves the system via Port 2 corresponding to the output of the circuit. The dispersion relation for SPPs propagation constants $\beta$ can be written in the form~\cite{teng}
\begin{equation}
\label{eq:1}
-k_{h}\left(\pm e^{-k_{h}d}-1\right)=2ik_{0}c{\varepsilon}_{d}{\varepsilon}_{0}/{\sigma}_{g},
\end{equation}
where $\sigma_{g}$ is the total conductivity of graphene, $k_{h}=\sqrt{{\beta}^{2}-k^{2}_{0}}$, $c$ is the speed of light in vacuum, $k_{0} =2\pi /\lambda_{0} $, $\varepsilon_{0} $ is the vacuum permittivity, $\varepsilon_{d}$ is the permittivity of the dielectric material between the graphene sheets with a distance $d$ between them. The pump SPP mode $E_{1}$ can also propagate in the graphene waveguide in Fig.~\ref{fig:1}a. Assuming that both SPPs interact with QD loaded in nanoresonator we will investigate the features of SPP mode patterns in waveguide for their simultaneous and separate propagation.

The surface conductivity of graphene $\sigma_{g} =\sigma_{\textrm{inter}} + \sigma_{\textrm{intra}} $ in the near-infrared range is given by the Kubo formula~\cite{falk} which consists of the intraband conductivity $\sigma_{\textrm{intra}} $ and interband conductivity $\sigma_{\textrm{inter}} $ and have the following forms:
\begin{subequations}
\label{eq:2}
\begin{eqnarray}
\sigma_{\textrm{intra}} &=& i\frac{8\sigma_{0} kT/h}{\omega +i/\tau } \left(\frac{\mu_{c}}{kT} +2\ln \left(e^{-\frac{\mu_{c}}{kT} } +1\right)\right), \\
\sigma_{\textrm{inter}} &=& i\frac{\sigma_{0}}{\pi} \ln \left(\frac{2\left|\mu_{c} \right|-\left(\omega +2i/\tau \right)\hbar}{2\left|\mu_{c} \right|+\left(\omega +2i/\tau \right)\hbar} \right),
\end{eqnarray}
\end{subequations}
where $k$ is the Boltzmann constant, $T$ is the temperature, $\mu_{c}$ is the chemical potential, $1/\tau$ is the carrier-scattering rate, $\sigma_{0} =\pi e^{2}/\left(2h\right)$ and $e$ is the electron charge, $h$ is the Planck constant. For simulation, we set the effective thickness of graphene as $\Delta_{g} =2 \; \textrm{nm}$, $\mu_{c} =0.6 \; \textrm{eV}$ (this value is typical for doped graphene), $\tau =0.9 \; \textrm{ps}$, $T=300 \; \textrm{K}$, wavelength of signal SPP mode $E_{2}$ $\lambda_{2} =8.04 \; \textrm{\textmu m}$, which configures the device to use it in infrared data transmission and processing systems.
\begin{figure}
\includegraphics[width=\columnwidth]{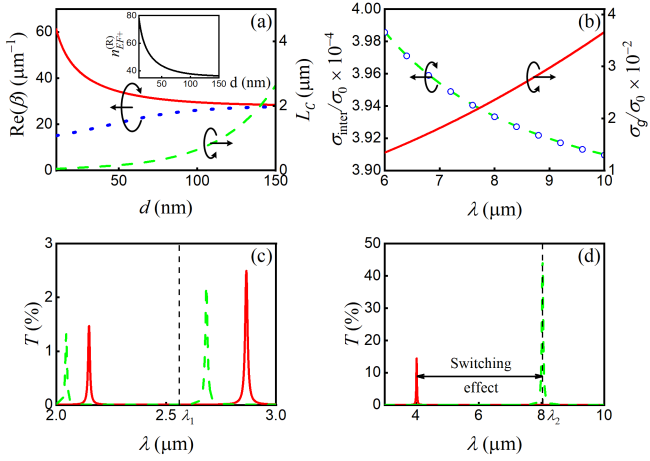}
\caption{\label{fig:2} (a) Propagation constants $\beta_{+}$ (solid red line) and $\beta_{-}$ (dotted blue line) together with the coupling length of signal SPPs $L_{C}$ (dashed green line) in the Port 1 for model in Figure~\ref{fig:1} versus the distance between graphene sheets; (b) Real part of total $\sigma_{g}$ (solid red line) conductivity and, separately, interband $\sigma_{\textrm{inter}}$ conductivity, calculated by using exact formula (\ref{eq:2}) (dashed green line) and Pade fit~\cite{pade} (circled blue line) versus free-space wavelength; (c) the transmittance of the stub resonator in vicinity of $\lambda_{1}$ in the absence (solid red line) and in the presence (dashed green line) of signal SPP for the set of coefficients $\left\{u\right\}=\left(t_{1},s_{1},s_{3},r_{3}\right)$, where $\left\{u\right\}=\left(0.1,0.9,0.065,0.9\right)$; (d) the transmittance of the stub resonator with InAs/ZnS QD in vicinity of $\lambda_{2}$ in the absence (solid red line) and in the presence of pump SPP mode $E_{1}$ (dashed green line) and $\left\{u\right\}$. The parameters $\sigma_{g}$ and $\sigma_{\textrm{inter}}$ are normalized to $\sigma_{0}$. In the inset for (a): the effective refractive index of $\beta_{+}$ versus intersheets distance.}
\end{figure}

\begin{figure}
\includegraphics[width=\columnwidth]{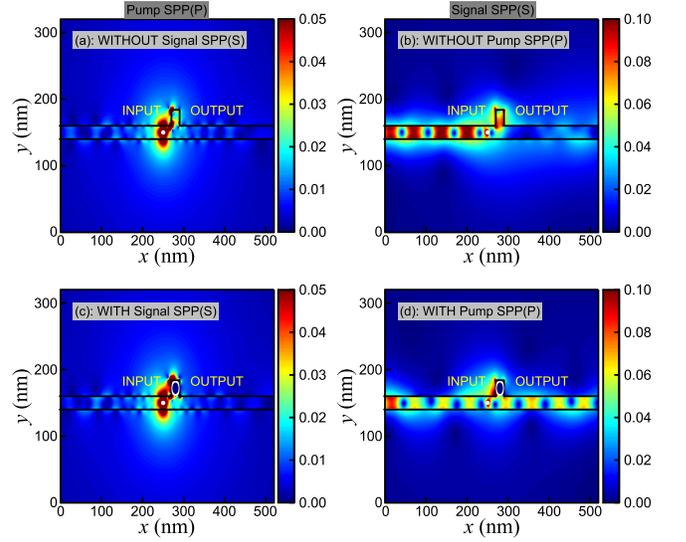}
\caption{\label{fig:fdtd} The summarized electric field $\sqrt{E_{x}^{2}+E_{y}^{2}}$ distributions (arbitrary units) for pump SPP (a), (c) and signal SPP (b), (d) in the stub nanoresonator loaded with QD. The switching between regimes of locking (b) and transmitting (d) of signal SPP is demonstrated. The black lines correspond to the graphene waveguide with stub nanoresonator and circled white line depicts the QD.}
\end{figure}

The propagation constants of the symmetric $\beta_{+}$ and anti-symmetric $\beta_{-}$ $E_{2}$ modes as a function of the distance between graphene sheets are presented in Fig.~\ref{fig:2}a. The space between sheets is filled with a dielectric $\varepsilon_{d} =2.022$. We will discuss the symmetric mode only because it leads to the highest density of electromagnetic field in the space between sheets. It is necessary to increase the efficiency of matter-field interaction with chromophore loaded in the stub nanoresonator~\cite{teng}. We choose $d=20 \; \textrm{nm}$, which determines a regime of strong coupling between sheets satisfying to condition $d<\xi$~\cite{wang1}, where $\xi =\textrm{Re}\left(\frac{\sigma_{\textrm{intra}}}{ic\varepsilon_0\varepsilon_d k_0}\right)=71 \; \textrm{nm}$. Based on the solution of equation (\ref{eq:1}), the estimations can be given for the other main characteristics of SPP propagation in the graphene waveguide, including SPP wavelength ${\lambda}_{SPP\pm}=\frac{2\pi}{\textrm{Re}\left({\beta}_{\pm}\right)}$, coupling length $L_{C}=\frac{\pi }{2\sqrt{2}\left|C_{g}\right|}$ and propagation length $\bar{L}_{SPP\pm}=\frac{{\lambda}_0}{4\pi \textrm{Im}\left(n_{EF\pm}\right)}$. Here $C_{g}=\frac{\beta_{-}-\beta_{+}}{2}$ is the coupling constant of SPPs and $n_{EF\pm} =n_{EF\pm}^{\left(\textrm{R}\right)}+in_{EF\pm}^{\left(\textrm{I}\right)}=\beta_{\pm}/k_{0}$ is the effective refractive index.
In our case, we have $n_{EF+} =59.3+0.17i$, ${\lambda}_{SPP+}=135.5 \; \textrm{nm}$, $L_{C}=74 \; \textrm{nm}$, $\bar{L}_{SPP+}=3.7 \; \textrm{\textmu m}$ for signal SPP. These estimations allow to achieve
about $150$ stub-nanoresonators at the distance equals to propagation length. The obtained values of the main parameters are in good agreement with the result of full-wave electromagnetic simulation based on FDTD (Finite Difference Time Domain) method~\cite{Sullivan}. We realized FDTD method by ourselves (see Fig.~\ref{fig:fdtd}) and tested it using other simulations for graphene~\cite{teng,Sarker,Hossain}.

We start with tuning the parameters of the passive stub so that in the absence of SPP mode $E_{1}$ it does not transmit SPP mode $E_{2}$ to the right side of the stub (Port 2). We optimized the height $D$ of stub resonator to achieve a transmittance minimum~\cite{stub1} at the signal wavelength $\lambda_{2}$ for a given stub width $d$. Since the phase shift in such a resonator is determined by the expression
\begin{equation}
\label{eq:4}
\phi\left(\lambda\right)=\frac{2\pi}{\lambda} \left(2D+d\right) n_{EF+}^{\left(\textrm{R}\right)}+\Delta \phi,
\end{equation}
then from condition $\phi\left(\lambda_{2}\right)=\left(2m+1\right)\pi $ we obtain $D=23.8 \; \textrm{nm}$ for $m=0$. Here $\Delta \phi$ is the additional phase shift results from the material properties inside the stub resonator. We can neglect it owing to the absence of pump SPP, because signal SPP doesn't interact with QD in the stub separately from the pump SPP. Then, the signal SPP at the wavelength $\lambda_{2}$ has a destructive interference and locked by the stub, see Fig.~\ref{fig:fdtd}b.

The dependencies of the transmittance coefficient $$T\left(\lambda \right)=\left|t_{1} +\frac{s_{1} s_{3}}{1-r_{3} e^{i\phi \left(\lambda \right)}} e^{i\phi \left(\lambda \right)}\right|^{2}$$ are shown in Fig.~\ref{fig:2}c,d, where parameters $r_{i}$, $t_{i}$, $s_{i}$ corresponds to the reflection, transmission and splitting coefficients in the $i$th cross-section ($i$th Ports) of the stub from Fig.~\ref{fig:1}a. These coefficients can be physically explained from the scattering matrix theory~\cite{Haus} and expressed in terms of characteristic impedances of media, surface modification and geometry of the system. In addition, they can be directly extracted using FDTD method~\cite{fdtd}. By varying the material parameters and profile of the substrate for the  graphene sheet~\cite{kong,Gu}, it is possible to achieve very narrow transmittance peaks at the selected wavelength, see Fig.~\ref{fig:2}c,d. However, using our tunings the transmittance for signal SPP is almost zero at $8 \; \textrm{\textmu m}$ and the SPP does not propagate to the right side of the stub (see Fig.~\ref{fig:fdtd}b).

We will now consider the possibility to control the SPP propagation due to plasmonic resonance with nanostructures~\cite{Chen}. Therefore, we proceed to consider the matter-wave interaction between QD and SPP in the graphene stub nanoresonator loaded with core-shell QD. Using the semi-classical approximation we describe the Ladder-type SPP-QD interaction, as well as within the density matrix formalism we describe the three-level scheme of electron/hole sublevels in QD, see Fig.~\ref{fig:1}b. The Hamiltonian of the system QD+SPPs has the following form:
\begin{subequations}
\label{eq:ham}
\begin{eqnarray}
H &=& H_{0}+H_{v}, \\
H_{0} &=& \hbar \left(\omega_{12}\left|2\right\rangle \left\langle 2\right|+\left(\omega_{12}+\omega_{23}\right)\left|3\right\rangle \left\langle 3\right|\right), \\
H_{v} &=& -\hbar \left(\widetilde{\Omega}_1\left|2\right\rangle \left\langle 1\right|+\widetilde{\Omega}^*_1\left|1\right\rangle \left\langle 2\right|+\widetilde{\Omega}_2\left|3\right\rangle \left\langle 2\right|+\widetilde{\Omega}^*_2\left|2\right\rangle \left\langle 3\right|\right),
\end{eqnarray}
\end{subequations}
where $H_{0}$ is the Hamiltonian of unexcited QD and $H_{v}$ is the Hamiltonian of SPP-QD interaction, $\widetilde{\Omega}_{1}=\Omega_{1}e^{i\omega_{1}t}$ and $\widetilde{\Omega}_{2}=\Omega_{2}e^{i\omega_{2}t}$ are the Rabi frequencies of pump and signal SPPs, respectively, $\Omega_{1}$ and $\Omega_{2}$ are the slowly varying amplitudes of Rabi frequencies, $\omega_{1}$ and $\omega_{2}$ are the frequencies of pump and signal SPPs, respectively, $\omega_{12}$ and $\omega_{23}$ are the frequencies of transitions in QD. Here the pump SPP mode $E_{1}$ is tuned to the interband transition $1S\left(h\right) \rightarrow 1S\left(e\right)$ while the signal SPP mode $E_{2}$ is tuned to the intraband transition $1S\left(e\right) \rightarrow 1P\left(e\right)$. We note that in the absence of pump SPP levels $\left|2\right\rangle$ and $\left|3\right\rangle$ are empty and signal SPP doesn't interact with QD, therefore $\Delta\phi=0$.

From a mathematical point of view, the novelty of our approach is the exploitation of nonlinear regime of SPP-QD interaction corresponds to the two-quantum transitions between levels $\left|3\right\rangle$ and $\left|1\right\rangle$ in the case $\Delta \gg \gamma_{31}, \; \Omega_{2}$, where $\Delta$ is the frequency detuning (see Fig.~\ref{fig:1}b) and $\gamma_{31}$ is the rate of decay between levels $\left|3\right\rangle$ and $\left|1\right\rangle$~\cite{our,our3}. Note, that the relaxation rates of QD have a dramatic increase only at the certain distance to the conductive mirror~\cite{Larkin}, but there exist optimized distances with maximum efficiency of energy conversion from QD into SPP~\cite{our4}. Then, under the condition $\Omega_{1}>\Omega_{2}$, the SPP mode $E_{1}$ provides the effective non-linear phase modulation of signal SPP. The key idea is to tune the pump SPP parameters so that the phase shift $\Delta\phi$ of signal SPP will be $\pi$ radians. This is the basis for realization of the switching effect and creation of all-plasmonic transistor.

Based on the known material parameters of graphene and $\textrm{A}_{\textrm{III}}\textrm{B}_{\textrm{V}}$ semiconductors we use the core-shell InAs/ZnS QD~\cite{Bouarissa,Cao1} to achieve our goal, see. Fig.~\ref{fig:1}a. The resonant frequencies of the corresponding transitions can be obtained in the form
\begin{subequations}
\label{eq:5}
\begin{eqnarray}
\omega_{12} &=&\frac{e E_{g}}{\hbar} +\frac{2 \hbar \kappa_{1,0}}{D_{QD}^{2}} \left(\frac{1}{m_{c}} +\frac{1}{m_{h}} \right), \\
\omega_{23} &=&\frac{2 \hbar}{D_{QD}^{2} m_{c}} \left(\kappa_{1,1}^{2} -\kappa_{1,0}^{2} \right),
\end{eqnarray}
\end{subequations}
where $E_{g} =0.35 \; \textrm{eV}$ is the band gap of InAs, $m_{c} =0.026m_{0}$, $m_{h} =0.41m_{0}$ are the effective masses of electron and hole, respectively; $\kappa_{1,1} =4.493$ and $\kappa_{1,0} =\pi$ are the roots of the Bessel function. The dipole moment of the intraband transition can approximately be estimated by the relation $\mu_{32} =0.433 e a_{QD} \Lambda$~\cite{Madelung}, where $\Lambda =3\varepsilon_{\textrm{ZnS}}/\left(2\varepsilon_{\textrm{ZnS}} +\varepsilon_{\textrm{InAs}}\right)$ and $a_{QD} =D_{QD}/2$ is the radius of the QD core with dielectric permittivity $\varepsilon_{\textrm{InAs}} =12.3$; the dielectric permittivity of the shell is $\varepsilon_{\textrm{ZnS}} =8.3$. The square of dipole moment of the interband transition can be found from Ref.~\citenum{mork}: $$\mu_{12}^{2} =\frac{e^{2}}{6m_{0} \omega_{1}^{2}} \left(\frac{m_{0}}{m_{c}} -1\right)\frac{E_{g} e \left(E_{g} +\Delta_{s} \right)}{E_{g} +2\Delta_{s}/3},$$ where $\Delta_{s} =0.43 \; \textrm{eV}$ is the spin-orbit splitting for InAs. All parameters of QD are determined by using our numerical simulator~\cite{plazm} and the initial data from the literature~\cite{Bouarissa,Cao1}.

We determine the QD size from (\ref{eq:5}b) based on the condition $\lambda_{23} \equiv 2\pi c/\omega_{23} =\lambda_{2}$ and obtain $a_{QD} =9.9 \; \textrm{nm}$. According to this QD core size and additional condition $\omega_{12} =\omega_{1}$ for pump SPP, we have determined resonance wavelength $\lambda_{1} =2.56 \; \textrm{\textmu m}$, as well as the values of the dipole moments $\mu_{12} =14.9 \times 10^{-29} \; \textrm{C m}$ and $\mu_{32} =5.91 \times 10^{-28} \; \textrm{C m}$ for interband and intraband transitions, respectively. We note that pump field $E_{1}$ is also localized on graphene waveguide (see Fig.~\ref{fig:fdtd}a,c), but pump SPP propagates in the weak coupling regime ($\xi=6 \; \textrm{nm}$). At the same time, the week coupling regime is compensated by a high intensity of the pump field $E_{1}$ at the input. The transmittance for pump SPP at the wavelength $\lambda_{1}$ is not so high for both cases with and without signal SPP, see Figs.~\ref{fig:2}c and~\ref{fig:fdtd}a,c.

The implementation of two SPPs $E_{1}$ and $E_{2}$ in the circuit leads to the appearance of polarization $\rho_{32}$ for transition $1S\left(e\right) \to 1P\left(e\right)$. We used the Liouville master equation with Hamiltonian (\ref{eq:ham}) and assumption of unchanged polarization $\rho_{31}$ (i.e. $\dot{\rho}_{31}=0$) to derive the nonlinear equation for dynamics of density matrix element $\rho_{32}$:
\begin{equation}
\label{eq:6}
\dot{\rho}_{32} =\sum_{i=1}^{4}X_{i},
\end{equation}
where $X_{1} =-i\Omega_{2} n_{32}$ corresponds to the induced single-quantum transitions in the system; $X_{2} =\Omega_{1}^{*} \Omega_{2} \rho_{21}/\left(i\delta +\gamma_{31} +\gamma_{32} \right)$ corresponds to the nonlinear cross-interaction between SPP modes, and $X_{3} =-\left|\Omega_{1}\right|^{2} \rho_{32}/\left(i\delta +\gamma_{31} +\gamma_{32} \right)$ corresponds to the nonlinear modulation of signal SPP induced by interaction with the pump field; $X_{4} =-\rho_{32} \left(\gamma_{21} +\gamma_{32} +\gamma_{31} +i\left(\delta -\Delta \right)\right)$ corresponds to the linear effects associated with the dispersion and spontaneous decay of the excited states; $\delta$ is the frequency detuning of signal SPP, see Fig.~\ref{fig:1}b. Now, we represent the Rabi frequencies in the form $\Omega_{1} =g_{1} B$ and $\Omega_{2} =g_{2} a$ for pump and signal SPP modes, respectively, with average numbers of plasmon-polaritons $\left|B\right|^{2}$ and $\left|a\right|^{2}$. The parameters $n_{ij} =\rho_{ii} -\rho_{jj}$ and $\gamma_{ij}$ determine the population imbalance and the rate of decay between levels $\left|i\right\rangle$ and $\left|j\right\rangle$, respectively.

\begin{figure}
\includegraphics[width=\columnwidth]{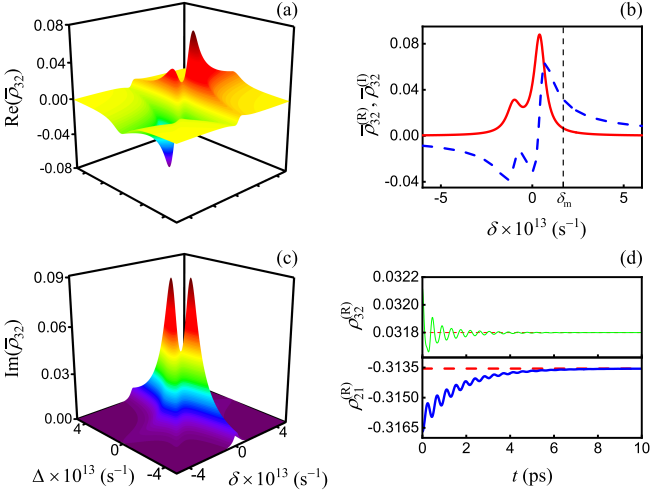}
\caption{\label{fig:4} The dependencies of the (a) real and (c) imaginary parts of $\bar{\rho}_{32}$ on the frequency detunings $\Delta$ and $\delta$; (b) the frequency dependencies of real (solid blue line) and imaginary (dashed red line) parts $\bar{\rho}_{32}$ for constant value of $\Delta_{m}$; (d) the time dependencies of the real parts of $\rho_{32}$ (thin green and red lines) and $\rho_{12}$ (thick blue and red lines), calculated by using formulas (\ref{eq:7}) (dashed lines) and by using direct numerical simulation (solid lines) of the full system of differential equations for density matrix elements upon Ladder-type interaction of two SPP modes and core-shell QD. The simulation parameters correspond to Fig.~\ref{fig:2}.}
\end{figure}

Applying the stationary conditions (i.e. $\dot{n}_{21}=\dot{n}_{32}=\dot{\rho}_{21}=\dot{\rho}_{31}=\dot{\rho}_{32}=0$) to our system allows to obtain the steady-state solutions for the polarizations (see Fig.~\ref{fig:4}d) in the following form~\cite{gau}:
\begin{subequations}
\label{eq:7}
\begin{eqnarray}
\bar{\rho}_{32} &=&-\frac{i\Omega_{2} \left(\left|\Omega_{1}\right|^{2} \bar{n}_{21} +D_{1} D_{2} \bar{n}_{32} +\left|\Omega_{2}\right|^{2} \bar{n}_{32}\right)}{\left|\Omega_{1}\right|^{2} D_{1} +D_{1} D_{2} \Gamma_{32} +\Omega_{2}^{2} \Gamma_{32}}, \\
\bar{\rho}_{21} &=&-\frac{i\Omega_{1} \left(\left|\Omega_{1}\right|^{2} \bar{n}_{21} +\left|\Omega_{2}\right|^{2} \bar{n}_{32} +D_{2} \bar{n}_{21} \Gamma_{32}\right)}{\left|\Omega_{1}\right|^{2} D_{1} +D_{1} D_{2} \Gamma_{32} +\left|\Omega_{2}\right|^{2} \Gamma_{32}},
\end{eqnarray}
\end{subequations}
where $D_{1} =i\Delta +\gamma_{21}$, $D_{2} =i\delta +\gamma_{31} +\gamma_{32}$, $\Gamma_{32} =i\left(\delta -\Delta \right)+\gamma_{21} +\gamma_{31} +\gamma_{32}$, $\bar{n}_{ij}$ is the stationary value of population imbalance between levels $\left|i\right\rangle$ and $\left|j\right\rangle$. The coupling SPP-QD constants have the form $$g_{1\left(2\right)} \left(\bar{r}\right)=\sqrt{\frac{\omega_{1\left(2\right)}}{\hbar \varepsilon_{0} V_{EF}}} \frac{E_{1\left(2\right)} \left(\bar{r}\right)}{E_{1\left(2\right)}^{\left(\max\right)} \left(\bar{r}\right)} \mu_{12\left(32\right)},$$ where the effective volume can be taken as $V_{EF} =\left(4/3\right)\pi a_{QD}^{3}$, and $E_{1\left(2\right)} \left(\bar{r}\right)$ is specified by the SPP mode distribution in the stub, and it can be obtained by the FDTD method~\cite{stub1}. The values of the coupling constants will be $g_{1} =6.575 \times 10^{11} \; \textrm{s}^{-\textrm{1}}$ and $g_{2} =1.472 \times 10^{12} \; \textrm{s}^{-\textrm{1}}$ for the selected parameters provided that the location of QD is in the geometrical center of the stub and taking into account the calculated intensities of signal and pump SPPs inside the stub, see. Fig.~\ref{fig:fdtd}.

In order for the transistor to be opened, the additional phase shift should be $\Delta \phi_{\max} =\pi \; \textrm{rad}$. In this case, the first order maximum of SPP mode $E_{2}$ will be precisely tuned to the wavelength of $8.04 \; \textrm{\textmu m}$, see Fig.~\ref{fig:2}d. In order to determine the required parameters of pump SPP mode $E_{1}$, we represent the expression for the phase shift in the form $\Delta \phi_{\max} =\left(2\pi/\lambda_{2}\right) n_{QD}^{\left(\textrm{R}\right)} D_{QD}$, where $n_{QD} =n_{QD}^{\left(\textrm{R}\right)} +in_{QD}^{\left(\textrm{I}\right)}$ is the complex nonlinear refractive index of QD, see Fig.~\ref{fig:4}a--c. The nonlinear refractive index $n_{QD} \approx \chi_{QD}/2$, where $\chi_{QD} =N \mu_{32} \bar{\rho}_{32}/\left(\varepsilon_{0} E_{2}\right)$, $N=5 \times 10^{19} \; \textrm{cm}^{-\textrm{3}} $ is the carrier concentration of InAs~\cite{Madelung}, $\bar{\rho}_{32} =\bar{\rho}_{32}^{\left(\textrm{R}\right)} +i\bar{\rho}_{32}^{\left(\textrm{I}\right)}$, $\bar{\rho}_{32}^{\left(\textrm{R}\right)} \equiv \textrm{Re}\left(\bar{\rho}_{32}\right)$, $\bar{\rho}_{32}^{\left(\textrm{I}\right)} \equiv \textrm{Im}\left(\bar{\rho}_{32}\right)$. Thus, we get the possibility to control the pattern of signal SPP spatial distribution inside and at the output of stub~\cite{evl2} (Port 3 in Fig.~\ref{fig:1}a) via parameters of the pump SPP mode.

As a signal mode, we choose a weak SPP with amplitude $a=1$, the field strength is represented by formula $E_{2} =\hbar \Omega_{2}/\mu_{32}=\hbar g_{2} a/\mu_{32}$ ($E_{1} =\hbar \Omega_{1}/\mu_{12}=\hbar g_{1} B/\mu_{12}$). The required phase shift $\Delta \phi_{\textrm{max}}$ is formed when the amplitude of the pump SPP mode is $B=10$. In this case, the destructive interference will change to constructive and the transistor will be opened for signal SPP propagating to the right side of the stub (compare regimes b and d in Fig.~\ref{fig:fdtd}). We note that the required regime can be achieved only under condition $\Delta > \gamma_{32}, \; \Omega_{2}$, when nonlinear interaction between SPPs and QD occurs most strongly. In the linear regime for cascade scheme with $\Delta=0$ the polarization $\textrm{Re}\left(\rho_{32}\right)$ does not achieve the necessary value against the background of large losses, see Fig.~\ref{fig:4}b.
In addition, we choose such combination of frequency detunings that the absorption $\alpha =N\mu_{32}^{2} \bar{\rho}_{32}^{\left(\textrm{I}\right)}/\left(2\textrm{Re}\left(\beta_{+}\right) \varepsilon_{0} \hbar g_{2}\right)$ of the signal SPP transmitted from Port 1 to Port 3 is kept low, see Fig.~\ref{fig:4}b,c. It takes place for the optimized values $\Delta_{m} =-6.156 \times 10^{12} \; \textrm{s}^{-\textrm{1}}$ and $\delta_{m} =1.697 \times 10^{13} \; \textrm{s}^{-\textrm{1}}$ and fixed values $\gamma_{21} =5 \times 10^{11} \; \textrm{s}^{-\textrm{1}}$, $\gamma_{32} =\gamma_{31} =1.43 \times 10^{12} \; \textrm{s}^{-\textrm{1}}$, that corresponds to strong SPP-QD coupling regime in nanoresonator loaded with QD. Hence, we obtain $\alpha =8.3 \times 10^{-9} \; \textrm{cm}^{-\textrm{1}}$ and $\bar{n}_{32} =-0.3429$, $\bar{n}_{21} =-0.3042$ (with stationary values of populations $\bar{\rho}_{11} =0.6504$, $\bar{\rho}_{22} =0.3462$, $\bar{\rho}_{33} =0.0034$).

The representation (\ref{eq:6}) for the polarization change rate allows us to understand the contribution of each linear (and nonlinear~\cite{Dzedolik1,Dzedolik2}) effects to the process of switching and achieving the stationary regime of SPP-QD interaction in the stub. In particular, the initial deviations for polarizations from the stationary values given by $\rho_{21\left(32\right)} \left(0\right)=1.01 \bar{\rho}_{21\left(32\right)}$ were used for simulation in Fig.~\ref{fig:4}d and led to the values of considered coefficients $X_{1} =5.05 \times 10^{11}i$, $X_{2} =-1.45 \times 10^{10} +1.78 \times 10^{11}i$, $X_{3} =-2.97 \times 10^{10} +7.68 \times 10^{10}i$, $X_{4} =4.42 \times 10^{10} -7.65 \times 10^{11}i$. The stabilization process takes place such that the formation of a strong phase shift resulting from $\textrm{Im}\left(X_{1}\right)$, $\textrm{Im}\left(X_{2}\right)$ and $\textrm{Im}\left(X_{4}\right)$ is compensated by nonlinear cross-interaction due to $\textrm{Im}\left(X_{3}\right)$ with opposite sign. On the other hand, the growth of linear losses resulting from $\textrm{Re}\left(X_{4}\right)$ is compensated by nonlinear induced amplification due to $\textrm{Re}\left(X_{3}\right)$. As a result, the system finds its own balance between dispersion, absorption and nonlinear processes and tends to a stationary level of parameters, which is a necessary condition for the stable functioning of the plasmonic device.

Now, let us to pay attention to the key working parameters of the designed device. Firstly, it is a size of transistor, which in our case is about $25 \; \textrm{nm}$. Secondly, a clock frequency that is about $100 \; \textrm{GHz}$ and can be obtained from the characteristic stabilization time of the system approximately $10 \; \textrm{ps}$, see Fig.~\ref{fig:4}d. And thirdly, a length of SPP propagation that is $3.7 \; \textrm{\textmu m}$ in our case allowing to place about 150 elements within the circuit.

The special attention should be paid to the production of the all-plasmonic transistor discussed here. The process of creating a device requires the combination of several experimental techniques at once. At the first stage, PECVD (plasma enhanced chemical vapor deposition) method should be used to deposit a high-quality graphene monolayer~\cite{Chun} on a silica substrate with a patterned stub.

Obviously, the stub corners cannot be right angles, as this will lead to the excessive stresses in graphene. The radius of curvature for the stub corners should be determined experimentally, but it must be more than $2 \; \textrm{nm}$ according to the sampling interval in our FDTD method. Note that in such conditions we observe a high density of near field in the corners of resonator (see Fig.~\ref{fig:fdtd}a,c), but this does not dramatically change the SPP propagation. At the second stage, the quantum dot need to be loaded into the stub using the nanomanipulation technique with atomic force microscope~\cite{Ratchford}. The next step is to coat the quantum dot and graphene sheet with dielectric by means of ALD (atomic layer deposition) method~\cite{Jeon,Ahn}. Finally, it is necessary to reapply PECVD method for deposition of second graphene layer on dielectric and ALD method for the final coating of device with a dielectric. We note that the field sources inside device can be implemented by core-shell nanowires~\cite{Ho} integrated inside plasmonic circuit.

\section{Conclusion}
In this paper we have considered the model of graphene stub nanoresonator loaded with a core-shell quantum dot to solve the problem of all-plasmonic control of SPP modes at the nanoscale. Based on the main parameters of graphene and InAs/ZnS semiconductors, we have investigated the physical properties, and their connections with geometric characteristics, of the stub nanoresonator and QD in order to realize switching between two stable states of a signal SPP mode using a pumping SPP. The strong coupling regime in the condition of Ladder-type nonlinear plasmon-exciton interaction scheme for QD loaded in the stub has been considered. As a result we have proposed the model of all-plasmonic transistor with $8 \; \textrm{\textmu m}$ working wavelength, $25 \; \textrm{nm}$ characteristic size and $100 \; \textrm{GHz}$ clock frequency. The proposed transistor can be used for development of a new perspective in processors architecture. In contrast to classical electronic or all-optical gates our system can provide the switching effect for SPP modes in near-infrared range at the nanoscale, accompanied by very narrow resonance lines.

For practical realization of a new transistor, the strict account for a charge carrier mobility~\cite{Kelardeh} and memory effect~\cite{Choi} in graphene are necessary. The solution of this problem requires the development of a new paradigm of material science directed on achievement of high-temperature superconductivity of graphene~\cite{sup1} and graphene-like materials~\cite{Kelardeh1}. Moreover it addresses to 2D stack-structures based on a combination of monolayers with different physical properties and QDs~\cite{Carmen,Praveena,brit,Kim}. The greatest interest here is associated with the realization of gates on the basis of graphene nanoresonators with QDs supporting magnetic dipole~\cite{Freitag,khoh}, quadrupole~\cite{Arnold1}, or anapole-like responses~\cite{bar}. Next development of the semiclassical~\cite{our1} and full-quantum electrodynamics theory~\cite{bon1,our} of near-field interactions in such systems should be continued as well.

\begin{acknowledgments}
A.V. Prokhorov thanks Prof. M.I. Stockman and Prof. A.B. Evlyukhin for helpful discussions. This work was supported by Foundation for Assistance to Small Innovative Enterprises (FASIE), Grant No. 2226GS1/37022 (START-1) and by Russian Foundation for Basic Research (project no. 17-42-330001\_r\_a).
\end{acknowledgments}

\bibliography{Shesterikov}

\end{document}